# THE SCALAR FIELD OF 5D GRAVITY AND THE HIGGS FIELD OF 4D PARTICLE PHYSICS: A POSSIBLE CONNECTION

Paul S. Wesson

Department of Physics and Astronomy, University of Waterloo, Waterloo, Ontario N2L 3G1, Canada

Abstract: The main results are reviewed of relating the scalar field of noncompactified Kaluza-Klein gravity to the Higgs field of particle physics. The embedding of 4D space-time in a 5D manifold can result in a variable cosmological 'constant' and particle masses tiny compared to the Planck value.

Addresses: Mail to Waterloo above; email: psw.papers@yahoo.ca

# THE SCALAR FIELD OF 5D GRAVITY AND THE HIGGS FIELD OF 4D PARTICLE PHYSICS: A POSSIBLE CONNECTION

1. Introduction

General relativity is an excellent theory of classical gravity, but has never really explained the nature and origin of its central character, mass. At the other end of the physics spectrum, quantum field theory attempts to give a local account of particle masses by the Higgs mechanism, which will be the main focus of the recently-completed Large Hadron Collider. If physics is an organic whole, there should be an intermediate semi-classical Higgs theory, which can reasonably be expected to throw light on both the classical and quantum domains. Such an account will be outlined here, using noncompactified Kaluza-Klein theory as a base, along with its implications for particle masses and cosmology.

This is a discussion paper. That is, it is a review with certain new results added, where it is possible to discern connections between the scalar field of 5D relativity and the Higgs field of 4D particle physics. The relations noted in Section 2 elucidate two problems, namely: disparate estimates of the cosmological 'constant' for large versus small systems; and the origin of particle masses which are tiny compared to the Planck value. (The relations of Section 2 are generic, and do not address specific data such as the ratio of vacuum to matter energy, which is observed to be approximately 0.74.) The mathematics necessary to back up the physics of the main text is given in the appendix. The subject involves input from particle physics [1-5], gravitation [6-15] and cosmology [16-20]. A review is available on the embedding of 4D physics in 5D manifolds [21], to which the present work may be regarded as an update. Novel ways of testing 5D theory are outlined in the conclusion.

2. A Semi-Classical Theory of Particle Mass

The obvious way to explain mass is to join and/or extend general relativity with the Standard Model of particle physics. There have been many attempts at this, though none has met with universal acceptance, and a new and hopefully better one is proposed below.

A widespread approach is to modify the action of general relativity, by taking the 4D Ricci or curvature scalar $^4R$ and multiplying it by $\phi^2$ where $\phi$ is a new scalar field. This Higgs field is the expectation value of a quantum field, and to account for the interactions of particles it is also common to add a term proportional to $\phi^4$ to the action [1-5]. The latter may be modified in other ways also, but the coupling for the gravitational sector of the theory is usually taken as still including the gravitational constant $G$. A consequence of this approach is that a cosmological constant appears with a magnitude proportional to the square of the Higgs field $\Lambda \sim \phi^2$.

The cosmological constant is central to all attempts to unify gravity with the interactions of particles. In straight Einstein general relativity, in a vacuum $^4R = 4\Lambda$, so $\Lambda$ is a measure of the 4D curvature or energy at a point in spacetime. It is also possible to regard $\Lambda$ as measuring the density and pressure of a kind of perfect fluid, with $\rho_v = -p_v/c^2 = \Lambda c^2/8\pi G$. In this way, when ordinary matter is present the total density and pressure are $\rho_t = \rho_m + \rho_v$, $p_t = p_m + p_v$ and the total source is described by an energy-momentum tensor $T_{\alpha\beta}$ which is balanced against the Einstein tensor for the fields $G_{\alpha\beta}$ with no explicit mention of $\Lambda$. However, this neat trick is somewhat deceptive, because the $c^2/8\pi G$ in the definition of the fluid properties $\rho_v$, $p_v$ is exactly cancelled by the $8\pi G/c^2$ in the coupling of the field equations. It is often more instructive to take Einstein's field equations in their original form, $G_{\alpha\beta} + \Lambda g_{\alpha\beta} = (8\pi G/c^2)T_{\alpha\beta}$. This form makes it clear that, because the covariant derivative of the metric tensor is zero, adding a term $\Lambda g_{\alpha\beta}$ to the field equations merely expresses a kind of gauge invariance. It expresses a freedom in gravitational theory similar to that of the addition of a scalar field in classical electromagnetism. This kind of freedom is indeed found in all equations of physics, where the laws are expressed as differential equations which are insensitive to constants that appear only on integration, where they are usually fixed by boundary conditions. Irrespective of how $\Lambda$ is viewed, it is not a trivial parameter, either in cosmology or particle physics. The universe appears to be $\Lambda$-dominated at the present time and was almost certainly so if there was an early, inflationary epoch. Models of particles frequently involve intense vacuum fields, which can be expressed by a $\Lambda$-like parameter, and which do not cancel themselves unless one invokes an overriding principle such as supersymmetry. A possible connection between these domains is provided by the suggestion that the big bang was really a quantum tunneling event in a $\Lambda$-dominated spacetime [4]. Unfortunately, this and other ideas are difficult to evaluate because there is disagreement about the size and sign of $\Lambda$.

The cosmological-constant problem is connected to the hierarchy problem, which in short is the relative weakness of gravity compared to the interactions of particles, or the observed smallness of particle masses compared to the theoretical Planck mass $(hc/G)^{1/2}$ of order $10^{-5}$g. It is reasonable to expect that the Higgs field – or something like it – would resolve both problems. Current models are inadequate, and the situation is compounded by the possibility of there being more than one Higgs field and uncertainties about the masses of the associated bosons [1,5]. In the basic theory of the electroweak interaction, the Higgs boson causes the *W* and *Z* gauge bosons to become massive, and also gives fermions their masses. But even in this simple version of the Higgs mechanism, a very heavy Higgs boson $(>10^3 GeV)$ would mean that the *W* and *Z* would become strongly interacting, making calculations nearly impossible. The Large Hadron Collider may provide a much-needed experimental perch in this subject, but doubts will probably remain about the theoretical status of the Higgs mechanism as it has been traditionally discussed.

Recent attempts to explain the nature of particle mass have focused on the idea that the world could have more than the 4 dimensions of spacetime. Specifically, 5D noncompactified general relativity has been the subject of intense study in recent years [6-15], and agrees with all extant observations [16-20]. It is similar to old Kaluza-Klein theory, but drops the restrictive 'cylinder' and 'compactification' conditions, yielding a rich and fully covariant algebra. This can be interpreted to provide new insights to what is commonly called rest mass and matter. In 5D membrane theory, all of the interactions are confined to a hypersurface which is identified with spacetime, except gravity which is allowed to propagate outside into the 'bulk' and thereby relatively weakened [3,21]. In 5D space-time-matter theory, all of the interactions are treated on the same basis, but the equations of motion mean that a particle moves only slowly away from a given hypersurface, with the variation of the potentials along the fifth dimension being responsible for (or inducing) matter in spacetime [6,21]. These versions of 5D relativity are actually the same from a mathematical standpoint, but differ in physical interpretation. It is not really necessary to make a choice about the philosophy behind these approaches, because both admit of a physical condition which enormously simplifies their consequences: It is known that a massive particle moving on a timelike path in 4D may be moving on a *null* path in 5D [11,12]. (That is, in terms of the usual 4D proper time and the equivalent 5D quantity, $ds^2 \geq 0$ corresponds to $dS^2 = 0$.) Another way of stating this is that it is possible to regard all particles in 4D – massive or not – as resembling photons in 5D.

To appreciate why this idea is feasible, and to glimpse its implications, it is useful to briefly reconsider some properties of mechanics.

In quantum mechanics, the prime parameters of a test particle are its energy *E*, its momentum *p*, and its rest mass *m*. Of course, the first pair of parameters intrinsically depend on *m*, and it is conventional to define a 4-vector $p^\alpha \equiv mu^\alpha$ where $u^\alpha \equiv dx^\alpha/ds$ are the 4-velocities. Locally in the weak-field limit, the geometry of spacetime is taken to be given by the Minkowski metric $M_4$, where the line element or proper time is $ds^2 = \eta_{\alpha\beta}dx^\alpha dx^\beta$ ($\eta_{\alpha\beta}$ = diagonal +1, – 1, – 1, – 1). The 4-velocities are conventionally normalized via $u^\alpha u_\alpha = 1$ for a massive particle, which thereby obeys the standard condition $E^2 - p^2c^2 = m^2c^4$. This expression is so ubiquitous in particle physics that it is often read in reverse as a *definition* for the mass $m$, assuming measurements are available for $E$ and $p$. However, it is important to realize that it is based on the normalization condition $u^\alpha u_\alpha = 1$, which does not contain *any* information about the nature of $m$, and in particular about the possibility that the mass varies in spacetime via $m = m(x^\alpha)$. Indeed, one can imagine that the mass depends on properties of a higher-dimensional manifold, such as the 5D coordinates $x^A$ ($A$ = 0, 123, 4 for time, ordinary space and the extra dimension); and then the regular 4D normalization condition $u^\alpha u_\alpha = 1$ when multiplied by the square of the 5D-dependent mass $m(x^A)$ will give exactly the *same* energy expression as before. This is not a trivial situation, either for experiment or theory. Even in Newtonian mechanics, a rocket accelerates because it burns fuel and decreases its mass. And in the mechanics of the early universe, particles are believed to have acquired mass through the agency of a scalar field, such as the Higgs field outlined above. It should also be noted that the energy relation noted previously can be regarded in the form $E^2 - p^2c^2 - m^2c^4 = 0$ as the result of the condition $u^A u_A = 0$ on the appropriately-defined 5-velocities. Equivalently, it is the result of $dS^2 = 0$ or the fact that the 5D interval is null.

In general relativity, the equations of motion for a test particle are commonly derived by extremizing the 4D interval. So a test particle follows the path with the least elapsed proper time. Since the interval is determined by the metric coefficients (potentials) and the coordinates, the equations of motion necessarily involve accelerations and velocities. Force is not a general-relativity concept. Newton realized that force is the rate of change of momentum (which involves the mass), and for a closed system it is the momentum which is conserved. Elementary quantum mechanics recognizes this, and its action is $I = \int mds$. It is obvious, therefore, that mass has to be introduced separately to relativity to match the dynamics of quantum theory. The two systems of dynamics are only equivalent when the mass is constant. This is usually the case in astrophysics, where masses change slowly if at all; but care is needed in applying metric-based dynamics to other systems. This is particularly true if the metric is extended to higher dimensions. In 4D, the normalization condition $u^\alpha u_\alpha = 1$ noted above implies when differentiated that the sum of the accelerations or forces per unit mass $f_\alpha$ obey the orthogonality relation $u^\alpha f_\alpha = 0$ $(\alpha = 0, 123)$. If an extra dimension is added, this changes to $u^A f_A = 0$ $(A = 0, 123, 4)$. Therefore, $u^\alpha f_\alpha = -u^4 f_4$ and what appears to be an extra force (per unit mass) generally appears. This has been isolated in both membrane theory and space-time-matter theory, and is presumably small [8-10]. However, momentum is observed to be conserved in 4D to high accuracy, and any 5 (or higher) D metric has to respect this. The lesson is that if a theory like general relativity is used as a template for higher-dimensional physics, the base metric has to be chosen carefully so as to incorporate mass and preserve the conservation of momentum.

The base metric of the world is widely assumed to be the Minkowski one, at least in 4D. However, $M_4$ is valid only locally in the weak-field limit. Globally, it is probably something else. If the present universe is dominated by dark energy described by a cosmological constant with $\Lambda > 0$, the base metric to a first approximation is likely de Sitter space. If the big bang was a quantum tunneling event, it likely involved $\Lambda < 0$ and anti-de Sitter space. If the dimensionality of the world is to be extended to 5D, it is not at all obvious that $M_5$ is the base space.

It is instructive, in identifying the base space of an $N$D theory, to consider a short historical / physical analogy. Imagine that an intelligent observer could walk around the Earth in Medieval times. With no more technology than a pot of paint, he could number prominent rocks which he passed in his peregrinations; and after sufficient time he would discover that he was bypassing some of the same rocks. He would conclude that he lived in a finite but unbounded 'space' of two spatial dimensions. Later in history, equipped with good surveying equipment, he would find that the world was, to a good approximation, the surface of a sphere. However, in order to do this he would have to formulate the concept of an extra dimension in his head, since his observational data would be restricted to those given by his feet. He would have no direct knowledge of the interior of the Earth, or the heavens above, and the extra dimension would *initially* be to him an abstraction. However, being intelligent, he would realize that for cartography it would be better to replace his measure of 2D distance $d\sigma^2 = dx^2 + dy^2$ by another one with a mysterious third axis, namely $d\sigma^2 = dr^2 + r^2\left(d\theta^2 + \sin^2\theta d\phi^2\right)$. In this, he could measure the angles $\theta$ and $\phi$ by means essentially the same as the ones used to determine latitude and longitude. But his 3D metric would still be a largely theoretical construct, because confined as he was to the Earth's surface he would have $dr = 0$ approximately, and accordingly he would set $r$ = constant = 1. The utility of his extra dimension would, however, become evident when later in history he gained the technological ability to bore beneath and rise above the planet's surface. Even excursions in *dr* of a few km, which were small compared to the Earth's radius of 6370 km, would show him that his extra dimension was not just a mathematical abstraction but a physical reality. Further, he would learn a lot about the physics of his world when he acquired a reasonably accurate gravimeter. For then, he would discover that the local acceleration *g* due to gravity was not exactly the same at all points on the Earth's surface. In actual fact, it is reported that *g* was considered a 'universal' constant in the era before Newton, who however in formulating $g = GM_E / R_E^2$ and relating it to the mass and radius of the Earth, set the scene for a better understanding of that parameter. Even so, the departures between the geometrical surface of the Earth and the surface of its gravitational potential (the geoid) are now known to be small. But they are finite. The situation in *N*D relativity at the present stage of the history of physics is similar to what has just been described. There are indications of the need for (at least one) extra dimension, and with it the need to redefine the base space of the world; but the accuracy of known physics implies that events happen close to the hypersurface of the higher-dimensional manifold we call spacetime, so new methods will probably be needed to convert the fifth dimension from an abstraction to a reality.

The analogy of the preceding paragraph has some practical implications. Let us assume that the world is described by a 5D Riemannian manifold, where the 4D interval $ds^2 = g_{\alpha\beta} dx^\alpha dx^\beta$ is replaced by the 5D one $dS^2 = g_{AB} dx^A dx^B$ (A, B = 0, 123, 4). Let the coordinates be chosen as usual to be the time $(x^0)$, the labels of ordinary space $(x^{123})$ and an extra one $(x^4 = l)$. The last is chosen so as to avoid confusion with the Cartesian measure and the implication that the fifth dimension is measured from some special hypersurface (as in membrane theory, where $y = 0$ is widely used to identify spacetime). Even in the case where an observer is confined exactly for some reason to the hypersurface $x^4 = l = $ constant, certain quantities which are connected with the fifth dimension will be present in the 4D hypersurface of spacetime. For example, $l$ itself and $dl/ds$, where we agree to use the 4D interval $s$ as a parameter in order to make contact with known physics. Quantities like these will have to be given some physical interpretation. Klein did this in 1926 when he interpreted the velocity in the extra dimension in terms of the electron charge $e$, in order to explain its quantization. That particular suggestion is now widely regarded as mistaken, and for reasons outlined above the present focus is not on electric charge but on particle rest mass $m$. However, the same principle applies. But in an unrestricted 5D manifold, the algebra is so rich in possibilities that the identification of the relevant functions in terms of measurable physical quantities is not easy. This applies not only to the metric and the equations of motion which flow from it, but also to the field equations and their associated physical quantities. The situation becomes even more complex in the case where an observer is allowed to move, even slowly, with respect to an $l$ - hypersurface (this is restricted by the fifth component of the 5D geodesic equation, which along with the field equations is discussed in mathematical terms in the appendix). To help cut down the complexity, it is helpful to neglect electromagnetic effects. Now in the old Kaluza-Klein theory, the electromagnetic potentials were identified with the extra components of the 5D metric $(g_{4\alpha}$ with $\alpha = 0,123)$. And the scalar poten-tial which is now of prime interest was artificially suppressed $\left(\left|g_{44}\right|=1\right)$. That old theory was a classical one describing a spin-2 graviton, a spin-1 photon and a spin-0 scaleron, all of which were massless [21]. The present philosophy is somewhat different. The scalar field finds a natural place as the fifth diagonal component of a 5D metric, superior to its inclusion as an *ad hoc* multiplier onto the 4D part of the metric (see above), and it should be retained as a possible analog of the Higgs field. However, the electromagnetic field can still be neglected, insofar as the current focus is on particle mass, and electromagnetic effects make only a small contribution to this. These and previous considerations are sufficient to enable a 5D metric to be written down which describes gravity plus a mass-related scalar field.

The desired 5D line element is given by

$$dS^2 = g_{\alpha\beta}\left(x^\gamma,\ell\right)dx^\alpha dx^\beta + \varepsilon\Phi^2\left(x^\gamma,\ell\right)dl^2 \ . \tag{1}$$

Here the scalar field $g_{44} = \varepsilon\Phi^2$ can take either sign, corresponding to a spacelike $(\varepsilon = -1)$ or timelike $(\varepsilon = +1)$ extra coordinate. Both are allowed by the field equations and a change can arise naturally if there is horizon-like behaviour in the extra dimension. If the signature is $(+---+)$, no trouble arises with closed timelike paths since the extra coordinate $x^4 = l$ does not have the physical properties of a time. Precisely what it means will become clearer below. For now, we remark that there is a large literature on metrics of form (1), including exact solutions of the field equations. The latter will be discussed shortly. Generically, one can add to (1) the suggestion outlined above, where conventional 4D causality $\left(ds^2 \geq 0\right)$ is included as the condition for a 5D null path $\left(dS^2 = 0\right)$. Also, one can redefine $\phi \equiv 1/\Phi$ (not to be confused with the spherical polar coordinate used before). Then (1) can be rewritten as $\left|dl^2\right| = \phi^2 ds^2$, which describes a conformal rescaling of spacetime by a scalar field. In fact, the 4D Ricci scalar always scales as $\phi^2$ for (1), as shown by equation (A6) of the appendix. This implies $\Lambda \sim \phi^2$, which is the same dependency of the cosmological constant on the scalar field as found in the standard approach to the Higgs mechanism.

The metric coefficients or potentials in the metric (1) have to be found by solving the field equations. The simplest choice for these is in terms of the 5D Ricci tensor:

$$R_{AB} = 0 \, (A, B = 0, 123, 4) \ . \tag{2}$$

This tensor also plays a central part in more complicated 5D theories, and from it can be formed other objects such as the 5D Einstein tensor $G_{AB}$, which contains the standard 4D one $G_{\alpha\beta}$. The 15 relations (2) break naturally into a set of 10 Einstein-like equations, a set of 4 Maxwell-like or conservation equations, and a single wave equation for the scalar field. These are given in full as equations (A3) – (A5) of the appendix, and discussed there. Below, attention will be focussed on the last. Before that, however, it is necessary to discuss the relationship between 5D and 4D geometrical quantities.

A 5D statement such as a set of field equations like (1) is independent of the choice of coordinates, or 5D covariant. Taking $R_{AB}$ as an example, it is always possible as an exercise in algebra to write it out at length, in terms of the metric coefficients and their first and second partial derivatives with respect to the coordinates. Some of these terms will be the ones found in the equivalent 4D expression, while others are new (these involve $g_{44}$ and the derivatives of the 4D metric coefficient $g_{\alpha\beta}$ with respect to the new coordinate $x^4 = l$). Again as an exercise in algebra, the 4D terms can be put on the left side of an equals sign and the new terms on the right side. So far, this is merely an example of the standard procedure in functional analysis. However, something significant is involved if the quantities concerned have a physical meaning. For there is now a relation which can be symbolically written as $Q$ (4D terms only) = $Q$ (5D-dependent terms). For the example of the Ricci tensor, its 4D components now satisfy the relation $R_{\alpha\beta}\left(g_{\alpha\beta}, x^\gamma \text{ only}\right) = R_{\alpha\beta}\left(g_{44}, \partial g_{\alpha\beta}/\partial l, \partial^2 g_{\alpha\beta}/\partial l^2\right)$. The left-hand-side is identical to the expression found in standard texts on general relativity; while the right-hand-side is an alternate expression, determined by how 4D is embedded in 5D. They may be called the intrinsic and extrinsic forms, since the one is determined entirely by operations in 4D spacetime while the other is determined by how the 4D metric is embedded in the 5D metric. There is sometimes confusion in the literature because authors fail to specify which form they are using, and occasionally both forms appear. It is important to realize that the forms are not contradictory, but *complementary*. For clarity, the statement $\Lambda \sim \phi^2$ made above comes from considering the extrinsic form for $^4R$. The splitting procedure outlined here can be applied to other objects, including $G_{AB}$, whose 4D part is $G_{\alpha\beta} = G_{\alpha\beta}\left(g_{\alpha\beta}, x^\gamma, g_{44}, l\right)$. Splitting this, and absorbing the constants for ease, gives a relation which can be written $G_{\alpha\beta} = 8\pi T_{\alpha\beta}$. Here $T_{\alpha\beta}$ is an effective energy-momentum tensor, dependent on the fifth dimension or induced by it. This interpretation of 5D relativity is the basis of space-time-matter theory. In it, Einstein's 4D field equations are satisfied by virtue of a 5D embedding. This approach is not, of course, the only possible one; but it does satisfy Einstein's dream of transmuting the "base-wood" of matter into the "marble" of geometry.

Embeddings are clearly of crucial importance for any 5D approach to the scalar field and its relation to particle mass and the Higgs mechanism. Considerable work was done on the subject in the 1990s from the physical side, before it was supported by the rediscovery of a relevant result on the mathematical side. Campbell's theorem is a relatively weak result of $N$D differential calculus, because it is restricted to local embeddings. However, it is of use for physics, because it guarantees that the 5D field equations $R_{AB} = 0$ contain the 4D Einstein equation $G_{\alpha\beta} = 8\pi T_{\alpha\beta}$ [13]. Another consequence of Campbell's theorem, which is also obvious in retrospect but seems odd initially, concerns the gauge-dependence of 4D physics in a 5D manifold. A 4D statement such as $G_{\alpha\beta} = 0$ is gauge-independent in the sense that it holds true under the group of 4D coordinate changes $x^\alpha \to x^\alpha(x^\beta)$. And a 5D statement like $R_{AB} = 0$ is gauge-independent under the change $x^A \to x^A(x^B)$. However, a 4D quantity $Q(x^\gamma, l)$ which depends on $x^4 = l$ will not in general keep its form under a change of coordinates that includes a change in $l$. That is, some 4D gravitational quantities are gauge-dependent and require the use of special gauges, in a manner reminiscent of the use of non-covariant gauges in quantum theory. This is why different physical interpretations are possible for a given 5D metric that satisfies the field equations [14,15]. Such cases occur frequently in the literature, but are primarily the result of using 4D thinking in a 5D covariant theory.

The choice of coordinates is important in 5D relativity in order to arrive at the appropriate physical interpretation of a given solution of the field equations. The subject is also more important than usually acknowledged in 4D relativity, where the Milne universe provides an instructive example. This has a Robertson-Walker line element, which in spatially isotropic coordinates is

$$ds^2 = c^2 dt^2 - \frac{R^2(t)}{\left(1 + kr^2/4\right)^2}\left[dr^2 + r^2 d\Omega^2\right] \quad . \tag{3}$$

Here $d\Omega^2 \equiv d\theta^2 + \sin^2\theta\, d\phi^2$, $R(t)$ is the scale factor and $k = \pm 1$ or $0$ is the 3D curvature constant. The Einstein field equation for (3) are the two Friedmann equations, which are satisfied for the Milne model with $R(t) = t$, $k = -1$, and no matter or vacuum sources $(\rho = p = 0 = \Lambda)$. The radial coordinate in (3) is chosen to be comoving with the overall expansion, so test particles are static in this frame and $r$ is merely a distance label. The 'real' distance at any time is proportional to $\int R(t) dr$. For the Milne case, the proper distance varies in proportion to $t$, which is a free expansion. Accordingly, one might suspect that (3) with $R \sim t$ and $k = -1$ is isometric to $M_4$. This is indeed so, with the coordinate transformation given in standard texts. Incidentally, all models with metric (3) can be embedded in 5D, and it is now known that they are all isometric to $M_5$ [7]. That is, the big-bang singularity can be viewed as the consequence of an unfortunate choice of coordinates.

The scalar field of 5D relativity is governed by the $R_{44} = 0$ component of the field equations (2). With metric (1), it reads

$$\Box \Phi = -\frac{\varepsilon}{2\Phi}\left[\frac{g^{\lambda\beta}{}_{,4}g_{\lambda\beta,4}}{2} + g^{\lambda\beta}g_{\lambda\beta,44} - \frac{\Phi_{,4}g^{\lambda\beta}g_{\lambda\beta,4}}{\Phi}\right]. \quad (4)$$

Here a comma denotes the ordinary partial derivative, and the semicolon in $\Box\Phi \equiv g^{\alpha\beta}\Phi_{,\alpha;\beta}$ denotes the ordinary 4D covariant derivative. This relation is interesting even from the purely algebraic standpoint, and deserves further study. Generally, it may be put into correspondence with the Klein-Gordon equation, provided an appropriate definition is made for the rest mass of a test particle in terms of derivatives of the potentials. Specifically, certain solutions of it are known which are relevant to the one-body and cosmological problems in 5D; but in general it is a wave equation with a source. The nature of the latter depends on coupling constants, and particularly on the sign of $\varepsilon(=\pm 1)$. This raises the notable possibility of determining the signature of the 5D manifold from the source. While (4) may in the appropriate limit be put into a form resembling Poisson's equation, the source is not in general the same as the ordinary matter which is the source for Einstein gravity (see the appendix). It is also possible that the right-hand side of (4) is zero, in which case it has the form of Laplace's equation. It should be recalled that to any solution of Poisson's equation may be added a solution of Laplace's equation, and that complicated solutions may be built up from simple ones by using Green's function. Above it was seen that $\Phi$ may be related to the Higgs field $(\phi \sim 1/\Phi)$. The inference is that solutions of (4) can provide a rich spectrum of possibilities for the Higgs field.

The canonical metric chooses the 5D coordinates in a particularly elegant manner, which leads to great simplification of the field equations and the equations of motion.

This is achieved by using the last available degree of coordinate freedom in the metric (1) to set the scalar field Φ to unity. Mathematically, this is acceptable, provided the 4D part of the metric $g_{\alpha\beta}(x^\gamma, l)$ is allowed to depend on the extra coordinate $x^4 = l$. Given this, it is also convenient to *l*-factorize the 4D metric, in a manner which has a particular application in mind. In membrane theory, this factor is usually chosen to be an exponential, giving the warp metric. In space-time-matter theory, it is usually chosen to be a quadratic, giving the canonical metric. Physically, factorizations of this kind have the inevitable consequence that some of the physics associated with the scalar field is compressed into the 4D sector, where it may be camouflaged by gravity. Nevertheless, both metrics have a large literature. The canonical case is especially easy to work with, when the extra dimension is taken to be spacelike and has a line element given by

$$dS^2 = (\ell/L)^2 g_{\alpha\beta}(x^\gamma, \ell) dx^\alpha dx^\beta - d\ell^2 \qquad (5)$$

$$= (\ell/L)^2 ds^2 - d\ell^2 \quad . \qquad (6)$$

Here $L$ is a constant length, introduced for the consistency of physical dimensions. Calling (5) $C_5$, the nature of $L$ may be determined by looking at the field equations for the pure-canonical case $C_5^*$ where $g_{\alpha\beta} = g_{\alpha\beta}(x^\gamma \text{ only})$. Then one recovers the Einstein equations with no matter but a cosmological constant given by $\Lambda = 3/L^2$. This is the intrinsic value of this parameter, determined by physics in the 4D hypersurface of spacetime (see above). The extrinsic value is $3/l^2$, which differs from the foregoing by the prefactor $l^2/L^2$ in (5) as expected. The sign of $\Lambda$ reverses when the extra dimension is taken to be timelike. Other properties of $C_5$ may be inferred by comparison with the 4D Milne universe discussed previously, as they are $l/t$ analogs in some regards. However, the 5D canonical metric $C_5$ and its pure form $C_5^*$ are not in general isometric to $M_5$. This can be appreciated by noting that the 4D Schwarzschild metric cannot be embedded in a flat manifold of less than 6 dimensions, whereas it can be embedded in $C_5^*$. With $\Lambda$ included, the 5D solution is

$$dS^2 = \frac{\Lambda l^2}{3}\left\{\left[1-\frac{2M}{r}-\frac{\Lambda r^2}{3}\right]dt^2 - \left[1-\frac{2M}{r}-\frac{\Lambda r^2}{3}\right]^{-1}dr^2 - r^2 d\Omega^2\right\} - dl^2 \quad . \tag{7}$$

The part of this metric inside the curly brackets is the 4D Schwarzschild-de Sitter one. Other 4D metrics may be similarly treated, and indeed it is a theorem that any vacuum solution of 4D general relativity may be embedded in the 5D pure-canonical metric. Also, the 4D equations of motion are reproduced *exactly* by the $C_5^*$ metric. (The subject of 5D dynamics is treated more fully in the appendix, but it is often convenient to short-circuit this by using the 5D null path and the metric directly.) It is interesting to note that in (7) and metrics like it, the role of $\Lambda$ is critical. In this regard, the shift $l \to (l-l_0)$ in metrics like (5) preserves the form of the metric but changes the cosmological "constant" to

$$\Lambda = \frac{3}{L^2}\left(\frac{l}{l-l_0}\right)^2 \quad . \tag{8}$$

This is only asymptotically equal to its previous intrinsic value of $3/L^2$ $(l \to \infty)$, and can diverge $(l \to l_0)$. This behaviour can be interpreted in different ways, depending on the physical meaning of $x^4 = l$ and the context [15]. In cosmology, the divergence can be iden-tified with the big bang, whereafter $\Lambda$ decays to an acceptable value as observed today. In particle physics, the divergence can be indentified with a resonance, wherein the vacuum field is locally intense. In either case, it is apparent that the cosmological-constant 'problem' can in principle be resolved, since $\Lambda$ is in general a variable in 5D theory.

More insight to the 5D theory is given by manipulating the field equations (see the appendix). It has been known for a while that when the 5D metric does not depend on the extra coordinate $x^4 = l$ in any way (not even via the quadratic factor $l^2/L^2$ of $C_5$), the trace of the effective energy-momentum tensor is $T = 0$, implying the equation of state $p = \rho/3$ typical of radiation. When the metric depends on $l$ only via the canonical quadratic factor (as in $C_5^*$), the effective equation of state is the $p = -\rho$ typical of the classical vacuum. Other dependencies on $l$ yield various kinds of 'ordinary' matter with their associated equations of state. However, it transpires that in general the source for the gravitational field ($g_{\alpha\beta}$) is *not* the same as the source for the scalar field ($\Phi$), though both involve matter. This leads to the suggestion that the conventional concept of a unique 'mass' as the source should be divided into two parts: the gravitational mass and the inertial mass. This is confirmed by the solitons, which comprise a class of exact solutions of the field equations (2) with metric

$$dS^2 = A^a dt^2 - A^{-a-b} dr^2 - A^{-a-b} r^2 d\Omega^2 \pm A^b dl^2 \quad . \tag{9}$$

Here $A \equiv (1 - 2M/r)$ where $M$ is a source at the centre of the 3D spherically-symmetric space that would conventionally be called the mass. The constants $a$, $b$ are related by the consistency relation $(a^2 + ab + b^2) = 1$. The properties of the solitons have been extensively studied [14, 21]. Since the metric coefficients in (9) do not depend on $x^4 = l$, the effective energy-momentum tensor given by (A3) of the appendix reads $8\pi T_{\alpha\beta} = \Phi_{,\alpha;\beta}/\Phi$, where $\Phi = A^{b/2}$. Accordingly, $T = 0$ and the equation of state is $p = \rho/3$. The appropriate interpretation is that a soliton is a centrally-condensed ball of radiation, consisting not of photons but of the massless scalerons of the $\Phi$ field. In the weak-field limit, the first and last parts of (9) lead to the view that the source has a gravitational mass $aM$ and an inertial mass $bM$. A more complete analysis, taking into account all of the metric coefficients, shows that the total energy of the soliton is $(a+b/2)M$. The possibility that mass is really bivalent, with gravitational and inertial aspects, needs more investigation. But it is clearly connected with the physical nature of the extra dimension.

The physical interpretation of the extra coordinate $x^4 = l$ of 5D relativity is controversial. This is unavoidable, since observation can only be compared to theory via coordinates, and in a 5D covariant theory all coordinates are in principle admissible. However, as in 4D theory, some coordinates are more convenient than others. There is a widespread feeling among workers that the rest mass $m$ of a particle is related to the coordinate $l$ or a function of it, in which case the scalar field $\Phi$ must be related to the Higgs field. A particularly strong case can be made for the identification $m \sim l/L$. The latter factor for $C_5$ gives back the standard element of 4D action $mds$ from (5), admits the conservation of ordinary momentum from the equations of motion, and agrees with the usual constant of the motion for energy in the appropriate limit; and also explains why the trace of the effective energy-momentum tensor is zero as for photons when the metric does not depend at all on $l$ (even via a quadratic factor). Fortunately, several of the proposed identifications for $m = m(l)$ become redundant if one adopts the hypothesis outlined above, that 4D causality with $ds^2 \geq 0$ is the consequence of a 5D null path with $dS^2 = 0$. For canonical coordinates with this condition, (6) gives $dl/ds = \pm l/L$, and the same relation emerges if the coordinate is inverted via $l \to L^2/l$. (So all of $m \sim l$, $m \sim 1/l$ and $m \sim dl/ds$ become essentially equivalent.) In this connection, it should be noted that when $\Phi$ is absorbed in $C_5$, the physics associated with the field is in a sense transferred to the coordinate itself. The condition $dS^2 = 0$ then necessarily results in an orbit in the $l/s$ plane. This orbit for $C_5^*$ can be either a growing mode or an oscillating mode, depending on the metric's signature. With a possible shift $l_0$ included, the two kinds of orbit are given in general by

$$l = l_0 + l_* e^{\pm s/L} \qquad (l = \text{spacelike}) \qquad (10)$$

$$l = l_0 + l_* e^{\pm is/L} \qquad (l = \text{timelike}) \quad . \qquad (11)$$

Here $l_*$ is a constant, which in the second mode is the amplitude of the oscillation. (In more general cases, $L$ in the above may be replaced by $L\Phi$ or $L/n$ where $n$ is the wave number.) For $C_5^*$ it should be recalled that $L = (3/|\Lambda|)^{1/2}$. Therefore, in this case, a test particle either wanders slowly away from a given $l$-hypersurface, or oscillates around it. In general, for a timelike extra dimension, the 4D metric is anti-de Sitter and the wave is supported by the vacuum.

Quantization of the 4D action can be obtained from the 5D null path for metrics of canonical type. However, by the previous discussion, it is to be expected that this process will be gauge-dependant and carried by the $l$ coordinate rather than the $\Phi$ field. This subject clearly needs a more detailed investigation, but to conclude this survey one can consider the relation $|dl/ds| = |l/L|$ which was noted above. This is generic, in that $l$ may be related to $m$ in different ways, notably ones which use gravitational units with $l = Gm/c^2$ (the Schwarzschild radius) and atomic units with $l = h/mc$ (the Compton wavelength). Choosing the latter, the usual rule for quantization of the 4D action follows from the noted 5D orbit. The relation concerned can be written most compactly in the form

$$\int \frac{ds}{l} = \int \frac{mcds}{h} = L \int \left| \frac{dl}{l^2} \right| = \frac{L}{l} = n \quad . \tag{12}$$

That is, the elementary rule for quantization is equivalent to the existence of periodic structure in the fifth dimension.

3. Conclusion

If particle rest mass is conferred by a Higgs field, it is highly desirable to unify this with the gravitational field of general relativity, to obtain a complete theory of the origin and effects of mass. This article has reviewed certain results which may be taken as indicating that there exists a semi-classical theory of the Higgs field which is compatible with Einstein gravity.

The problem of how to extend general relativity so as to account for particle masses is a difficult one. The present account has concentrated on the 5D approach, mainly because it offers the most natural home for a scalar field. Noncompactified, unrestrained 5D relativity is algebraically rich; but its covariance also means that there are several possible ways to identify particle mass. The trick is to put the 5D metric into a form where the 4D mass can be recognized, and thereby identified in terms of the scalar field and its associated extra coordinate. The canonical metric $C_5$ does this; and in its pure form $C_5^*$, where there is no intrusion of the extra dimension into spacetime, it provides a perfect embedding for general relativity, both in terms of the fields and the dynamics. Indeed, one can argue that the base metric of the world is $C_5$, not the $M_5$ commonly assumed. However, while $C_5$ is algebraically general, it suppresses the physics of the scalar field. To remedy this, more study is needed of the wave equation (4) which governs that field. This is justified, because its properties indicate that a range of solutions for it can be constructed, which together with an appropriate definition for mass (say $\Phi l$) would help to resolve the hierarchy problem. The cosmological-'constant' problem is also open to resolution, because in 5D that parameter is related to the scalar field $\left(\Lambda \sim 1/\Phi_{5D}^2 \sim \phi_{4D}^2\right)$ and is variable.

The Large Hadron Collider is expected to discover the Higgs boson associated with the scalar field and measure its mass (if it is not too large). There are, however, other ways to test an extra dimension which have hitherto not attracted much attention because they involve the electromagnetic sector of the theory [21]. For example, the Maxwell potentials as defined in 5D Kaluza-Klein theory actually involve the scalar field (see the appendix). This may be expected to show up as a possible variation in the permittivity of the vacuum, either in time or space. Such variability is expected to be small, but can in principle manifest itself by changes in the effective fine-structure constant for distant astronomical sources like quasars, or in changes to the short-range interactions and the $S$-matrix for scattered particles. These and other tests for an extra dimension have certain advantages: they can be concretely formulated and cheaply carried out.

Acknowledgements

This work was partially supported by N.S.E.R.C., and incorporates comments from members of the S.T.M. Consortium (http://astro.uwaterloo.ca/~wesson).

Appendix: The Field Equations and the Equations of Motion in 5D

There follow two paragraphs with comments, of a mathematical nature. Upper-case Latin (English) letters run 0,123,4 for the time $t$, space ($xyz$ or $r\theta\phi$) and the extra coordinate $x^4 = l$. Lower-case Greek letters run 0,123. Units are chosen so that the gravitational constant ($G$), Planck's constant of action ($h$) and the speed of light ($c$) are all unity. The coordinates are normally taken to have the physical dimensions of lengths while the metric coefficients in 5D and 4D $(g_{AB}, g_{\alpha\beta})$ are dimensionless, so the cosmological constant $\Lambda$ has physical dimensions of (length)$^{-2}$. A comma denotes the ordinary partial derivative, while a semicolon denotes the 4D covariant derivative.

The field equations in 5D are commonly taken in terms of the 5D Ricci tensor to be

$$R_{AB} = 0 \quad (A, B = 0,123,4) \ . \tag{A1}$$

The 15 components of this may be expanded, and grouped into sets of 10, 4 and 1 as noted below, once a form has been chosen for the 5D line element. For neutral matter, it is convenient to use 4 of the available 5 degrees of coordinate freedom to set the potentials of electromagnetic type to zero $(g_{\alpha 4} = 0)$. The remaining fifth degree of freedom is sometimes used to set the magnitude of the scalar potential to unity, but this may cause the physics associated with the fifth dimension to be 'compressed' into the 4D gravitational sector with $g_{\alpha\beta} = g_{\alpha\beta}(x^\gamma, l)$. This step will therefore be delayed until later, and for now the algebra proceeds with $g_{44} \equiv \varepsilon \Phi^2$, where $\Phi = \Phi(x^\alpha, l)$ and $\varepsilon = \pm 1$ allows for both a spacelike and timelike extra dimension. The 5D line element then takes the form

$$dS^2 = g_{\alpha\beta}\left(x^\gamma, \ell\right) dx^\alpha dx^\beta + \varepsilon \Phi^2\left(x^\gamma, \ell\right) dl^2 \quad . \tag{A2}$$

With this metric, the field equation (A1) can be instructively written as follows:

$$G_{\alpha\beta} = 8\pi T_{\alpha\beta}$$

$$8\pi T_{\alpha\beta} \equiv \frac{\Phi_{,\alpha;\beta}}{\Phi} - \frac{\varepsilon}{2\Phi^2}\left\{\frac{\Phi_{,4} g_{\alpha\beta,4}}{\Phi} - g_{\alpha\beta,44} + g^{\lambda\mu} g_{\alpha\lambda,4} g_{\beta\mu,4}\right.$$

$$\left. - \frac{g^{\mu\nu} g_{\mu\nu,4} g_{\alpha\beta,4}}{2} + \frac{g_{\alpha\beta}}{4}\left[g^{\mu\nu}{}_{,4} g_{\mu\nu,4} + \left(g^{\mu\nu} g_{\mu\nu,4}\right)^2\right]\right\} \quad . \tag{A3}$$

$$P^\beta_{\alpha;\beta} = 0$$

$$P^\beta_\alpha \equiv \frac{1}{2\Phi}\left(g^{\beta\sigma} g_{\sigma\alpha,4} - \delta^\beta_\alpha g^{\mu\nu} g_{\mu\nu,4}\right) \quad . \tag{A4}$$

$$\Box\Phi = -\frac{\varepsilon}{2\Phi}\left[\frac{g^{\lambda\beta}{}_{,4} g_{\lambda\beta,4}}{2} + g^{\lambda\beta} g_{\lambda\beta,44} - \frac{\Phi_{,4} g^{\lambda\beta} g_{\lambda\beta,4}}{\Phi}\right]$$

$$\Box\Phi \equiv g^{\alpha\beta} \Phi_{,\alpha;\beta} \quad . \tag{A5}$$

These sets of equations come respectively from $R_{AB} = 0$ when the latter is split into $R_{\alpha\beta} = 0$, $R_{\alpha 4} = 0$ and $R_{44} = 0$. The first set balances the Einstein tensor with an effective energy-momentum tensor which is induced from the fifth dimension. The $T_{\alpha\beta}$ of (A3) has been extensively studied. It has good algebraic properties and describes all known kinds of matter. A useful result of the algebra which leads to (A3) is a relation for the 4D Ricci or curvature scalar in terms of the embedding:

$$^4R = \frac{\varepsilon}{4\Phi^2}\left[g^{\mu\nu}{}_{,4} g_{\mu\nu,4} + \left(g^{\mu\nu} g_{\mu\nu,4}\right)^2\right] \quad . \tag{A6}$$

This is sometimes called the extrinsic form, as opposed to the intrinsic form, which is the standard one found in textbooks on general relativity and is defined entirely in terms of the 4D quantities of spacetime. Incidentally, the rule for converting from the extrinsic form of a scalar quantity to the intrinsic form for the canonical metric may be found by considering how the prefactor $(l/L)^2$ enters the 5D algebra, or by treating it as a conformal factor on the 4D part of the metric and using the standard transformation. The rule is that the extrinsic form times $(l/L)^2$ gives the intrinsic form, so for example the cosmological constant goes from $3/l^2$ to $3/L^2$ in applications of the first set of field equations (A3). The second set of field equations (A4) can be interpreted as conservation laws, and may be related to the 4 Maxwell equations, which must still be satisfied ( perhaps trivially) even though the electromagnetic-type potentials in the metric (A2) have been set to zero. The tensor $P_\alpha^\beta$ has an associated scalar $P = -3g^{\lambda\sigma}g_{\lambda\sigma,4}/2\Phi$. The physical dimensions of $P_\alpha^\beta$ are (length)$^{-1}$, so technically its 'square' can be regarded as a kind of source term similar to $T_\alpha^\beta$. This approach is sometimes used in membrane theory, but physically is questionable since $P_\alpha^\beta$ is conserved by itself as shown by (A4). The last of the 5D field equations (A5) has no counterpart in classical 4D theory, though its scalar potential $\Phi$ has occasionally been linked to the Higgs field of quantum 4D theory. In the 5D context, it obeys a wave equation whose source can be compared to the $T_{\alpha\beta}$ of (A3) which is the source for conventional gravity. In the case where $g_{\alpha\beta,44}$ dominates other terms, the sources for the gravitational and scalar fields are similar by (A3) and (A5), and in the weak-field limit both reduce to Poisson's equation. However, taking the trace of (A3) allows (A5) to be rewritten as

$$\frac{\Box \Phi}{\Phi} = 8\pi T + \frac{\varepsilon}{2\Phi^2} \left\{ \frac{\Phi_{,4} g^{\alpha\beta}}{\Phi} g_{\alpha\beta,4} - g^{\alpha\beta} g_{\alpha\beta,44} + \frac{\left(g^{\alpha\beta} g_{\alpha\beta,4}\right)^2}{2} \right\} \quad . \tag{A7}$$

This shows that in general the source for the scalar field is not the same as that for the gravitational field, though both involve 'ordinary' matter. (However, at a matter/vacuum interface where the discontinuity in $g_{\alpha\beta,44}$ can dominate other terms, $\Box\Phi/\Phi = -\varepsilon g^{\alpha\beta} g_{\alpha\beta,44}/2\Phi^2 = -8\pi T$ .) Another way to express (A7) is to use the scalar $P$ noted above which is associated with (A4). Then (A7) can be rewritten as

$$\varepsilon \Box \Phi = \frac{P_{,4}}{3} + \frac{g^{\alpha\beta}{}_{,4} g_{\alpha\beta,4}}{4\Phi} \quad . \tag{A8}$$

This shows that the scalar field is associated with the matter currents described by (A4).

The equations of motion in 5D are commonly derived by taking the extremum of the line interval, as given symbolically by $\delta\left[\int dS\right] = 0$. Alternatively, they may be obtained by using the Lagrangian, or by expanding the 5D geodesic equation. As with the field equations, the equations of motion will have a form dependent on the assumed form of the metric (though their results will be independent of the choice of coordinates since the theory is 5D covariant). If electromagnetic effects are included, the common form for the metric is

$$dS^2 = ds^2 + \varepsilon \Phi^2 \left(dx^4 + A_\mu dx^\mu\right)^2 \quad . \tag{A9}$$

Here the 5D line interval includes the 4D one which defines proper time, and it is often convenient to use *s* instead of *S* in order to make contact with extant knowledge. The 4D part of the 5D line element is given by $ds^2 = g_{\alpha\beta}(x^\gamma, l) dx^\alpha dx^\beta$, and the 4-velocities are $u^\alpha \equiv dx^\alpha / ds$ as usual. It should be noted that the 4D electromagnetic potentials in (A9) are defined in terms of the 5D metric by $A_\mu \equiv \varepsilon g_{4\mu} / \Phi^2$ and so depend intrinsically on the scalar field. The equations of motion for (A9) may be found in the literature, but are cumbersome [8, 21]. They show, however, that in the appropriate limit there is a constant of the motion associated with the zeroth or time component, given by $(g_{00})^{1/2}(1-v^2)^{-1/2}$ where *v* is the projected velocity in ordinary 3D space. This constant of the motion is conventionally identified with the energy of a test particle of rest mass *m*, which parameter may thereby be given a 5D interpretation if so desired. The equations of motion are also cumbersome for the metric (A2), even though it omits the electromagnetic potentials. Partly for this reason, it is convenient to use the fifth degree of coordinate freedom left in (A2) to set the magnitude of $g_{44} = \varepsilon\Phi^2$ to unity. Here it is usual to choose $\varepsilon = -1$ so the extra dimension is spacelike, though the timelike case can be determined from the following equations by a simple permutation. It is also convenient at this stage to factorize the 4D part of the 5D metric with a term in $x^4 = l$. This is commonly taken to be an exponential and gives the so-called warp metric, or a quadratic and gives the so-called canonical metric. The latter choice is especially effective in simplifying the field equations and the equations of motion. The metric is still algebraically general so long as the post-factorized 4D metric tensor is allowed to depend on *l*. (When it does not do so, the result is the so-called pure-canonical metric, which though is algebraically special.) The result of these choices is

$$dS^2 = (\ell/L)^2 g_{\alpha\beta}(x^\gamma, \ell) dx^\alpha dx^\beta - d\ell^2 \quad . \tag{A10}$$

Here $L$ is a constant length, which by reduction of the field equations from 5D to 4D can be identified when $g_{\alpha\beta} = g_{\alpha\beta}(x^\gamma$ only) in terms of the cosmological constant. The precise relation is $\Lambda = 3/L^2$. This changes sign when the last term in (A10) is swapped (by $l \to il$ with $L \to iL$). The equations of motion for (A10) are remarkably simple. They are best presented as a set for spacetime plus an extra relation for the motion in the fifth dimension:

$$\frac{d^2 x^\mu}{ds^2} + \Gamma^\mu_{\alpha\beta} \frac{dx^\alpha}{ds} \frac{dx^\beta}{ds} = f^\mu$$

$$f^\mu \equiv \left( -g^{\mu\alpha} + \frac{1}{2} \frac{dx^\mu}{ds} \frac{dx^\alpha}{ds} \right) \frac{d\ell}{ds} \frac{dx^\beta}{ds} \frac{\partial g_{\alpha\beta}}{\partial \ell} \tag{A11}$$

$$\frac{d^2 \ell}{ds^2} - \frac{2}{\ell} \left( \frac{d\ell}{ds} \right)^2 + \frac{\ell}{L^2} = -\frac{1}{2} \left[ \frac{\ell^2}{L^2} - \left( \frac{d\ell}{ds} \right)^2 \right] \frac{dx^\alpha}{ds} \frac{dx^\beta}{ds} \frac{\partial g_{\alpha\beta}}{\partial \ell} \quad . \tag{A12}$$

These relations show that the motion in spacetime is the standard geodesic one (where the Christoffel symbol accounts for the 4D curvature) but modified by a fifth force $f^\mu$, which is really an acceleration per unit (rest) mass. It is inertial in the Einstein sense, because it is proportional to the relative velocity between the 4D and 5D frames and their coupling (via $dl/ds$ and $\partial g_{\alpha\beta}/\partial l$). It is in general finite, but is zero for the pure-canonical case $\left( \partial g_{\alpha\beta}/\partial l = 0 \right)$. Now it is a consequence of Campbell's theorem that any vacuum solu-tion of the Einstein field equations $G_{\alpha\beta} + \Lambda g_{\alpha\beta} = 8\pi T_{\alpha\beta} = 0$ can be embedded in the pure-canonical metric. This includes the Schwarzschild solution for the solar system and the de Sitter solution for inflationary cosmology. (Other cosmological solutions with ordinary matter require $\partial g_{\alpha\beta} / \partial l \neq 0$, but to a first approximation the universe appears to be $\Lambda$-dominated: see refs. 16-20). It follows that the best data available cannot distinguish between 4D and 5D, at least by dynamical means.